\documentstyle[aps,prl]{revtex}         % For Galley Format
\begin{document}

\draft

\title{ 
Renormalization of the Inverse
Square Potential}

\author{Horacio E. Camblong,$^{1}$
Luis N. Epele,$^2$ Huner Fanchiotti,$^2$
 and Carlos A. Garc\'{\i}a Canal$^2$}

\address{$^1$ Department of Physics, University of San Francisco, San
Francisco, California 94117-1080 \\
$^2$  Laboratorio de F\'{\i}sica Te\'{o}rica,
 Departamento de F\'{\i}sica,
Universidad Nacional de La Plata,
\\
 C.C. 67 -- 1900 La Plata, Argentina}

\maketitle

\begin{abstract}
The quantum-mechanical 
$D$-dimensional inverse square potential
is analyzed using field-theoretic renormalization techniques.
A solution is presented for both the bound-state
and scattering sectors of the theory
using cutoff and dimensional regularization.
In the renormalized version of the theory, there is 
a strong-coupling regime where quantum-mechanical breaking
of scale symmetry takes place through dimensional transmutation, 
with the creation of a single bound state and of an
energy-dependent $s$-wave scattering matrix element.
\end{abstract}


%\narrowtext

The quantum-mechanical inverse square potential 
is a singular problem that has generated controversy for decades.
For instance, the solution proposed in Ref.~\cite{cas:50}
failed to give a Hamiltonian bounded from below,
and this led to a number of alternative
regularization techniques~\cite{fra:71,par:73,sch:76}
based on appropriate parametrizations of the potential---including the
replacement~\cite{all:72} of self-adjointness by an
interpretation of the ``fall of the particle to the 
center''~\cite{lan:77}.
However, it is generally recognized that the singular nature of this 
problem lies in that its Hamiltonian, 
being symmetric but not self-adjoint,
admits self-adjoint  extensions~\cite{sim:70}.
Recently, a renormalized solution was presented 
using field-theoretic techniques~\cite{gup:93},
but it was just limited to the one-dimensional case and 
cutoff renormalization.

In this Letter (i) we generalize the results of
Ref.~\cite{gup:93} to $D$ dimensions
(including the all-important $D=3$ case)
using cutoff regularization in configuration
space; (ii) present a complete picture 
of the renormalized theory; and (iii) confirm the
same conclusions using dimensional regularization~\cite{bol:72}.
This problem is crucial for the 
analysis and interpretation of the 
point dipole interaction of molecular physics~\cite{lev:67,des:94}, 
and may be relevant in polymer physics~\cite{mar:91}.
In addition 
(i) it displays remarkable similarities with the
two-dimensional $\delta$-function 
potential~\cite{tho:79,adh:95a,man:94};
(ii) it provides another
 example of dimensional transmutation~\cite{col:73}
in a system with a finite number of degrees of freedom;
and (iii) it illustrates the relevance of field-theoretic
concepts in quantum mechanics~\cite{tho:79,adh:95a,man:94,adh:95b}.

This problem is ideally suited for
implementation in configuration space~\cite{ara:97}, where the 
radial Schr\"{o}dinger equation
for a particle subject to the  $r^{-2}$ potential in
$D$ dimensions~\cite{lou:60} reads
(with $\hbar=1$ and $2 m=1$)
\begin{equation}
\left[
\frac{1}{r^{D-1}} \frac{d}{dr} \left( r^{D-1} \frac{d}{dr} \right)
\,
+ E - 
\frac{ l ( l + D -2 ) - \lambda }{r^{2}}
\right]  R_{l}(r) = 0
\;  ,
\label{eq:ISP_radial}
\end{equation}
which is explicitly scale-invariant  
because $\lambda$ is dimensionless~\cite{jac:72}.
In Eq.~(\ref{eq:ISP_radial}),
$l$ is the angular momentum quantum number and
$\lambda>0$ corresponds to an attractive potential;
with the transformation 
$
R_{l}(r)= r^{-(D-1)/2}\, u_{l}(r)
$, Eq.~(\ref{eq:ISP_radial}) is recognized to have
solutions
$
  R_{l}(r) 
=
r^{-(D/2 -1)}
\,
Z_{s_{l}}
(\sqrt{E} \, r
)
$,
where $Z_{s_{l}}(z) $ represents an appropriate linear combination
of Bessel functions of order 
$
s_{l}
=
[  \lambda_{l}^{(\ast)}
- \lambda
]^{1/2}  
$,
with
\begin{equation}
\lambda_{l}^{(\ast)} 
= ( l +  D/2 -1  )^{2}
\;  .
\label{eq:ISP_critical_coupling}
\end{equation}
If $\lambda$ were allowed to vary,
one would see that the nature of the solutions changes
around the critical value  
$\lambda_{l}^{(\ast)} $, for each angular momentum 
state.
For $ \lambda <\lambda_{l}^{(\ast)} $ (including repulsive potentials),
the order $s_{l}$ of the Bessel functions  is real,
so that the solution regular at the origin is proportional to 
the Bessel function of the first kind $J_{s_{l}}
\left(\sqrt{E} \, r
\right)$. However, the same solution fails to satisfy the 
required behavior at infinity for bound states ($E<0$);
in other words, in the weak-coupling regime,
the potential cannot sustain bound states.
Moreover, the scattering solutions 
are scale-invariant~\cite{jac:72},
with $D$-dimensional phase shifts
$\delta_{l}^{(D)} =
\{  [ \lambda_{l}^{(\ast)} ]^{1/2}
-
[ \lambda_{l}^{(\ast)} -\lambda ]^{1/2}  \} \pi/2$.
Nothing is surprising here: the potential $r^{-2}$ is
explicitly scale-invariant and
no additional scale arises at the level of the solutions,
which are well-behaved---one could say that the potential
looks like a regular ``repulsive'' one.
However, this picture changes dramatically
for $ \lambda >\lambda_{l}^{(\ast)} $:
 all the Bessel functions acquire
an uncontrollable oscillatory character
through the imaginary order $s_{l}=i \Theta_{l}$,
where
$
\Theta_{l} 
= [ \lambda - \lambda_{l}^{(\ast)} ]^{1/2}
$, as we shall see next. 

For the remainder of this Letter,
we will mainly analyze the strong-coupling 
regime  $ \lambda >\lambda_{l}^{(\ast)} $.
First, for the bound-state sector,
from Eq.~(\ref{eq:ISP_radial}),
$u_{l}(r)\propto \sqrt{r} \, K_{i\Theta_{l}} (\sqrt{|E|}\, r)$,
with $K_{s_{l}} (z)$ 
being the modified Bessel function of the second kind~\cite{abr:72},
whose behavior near the origin is of the form
\begin{equation}
K_{i\Theta_{l}} (z) 
  \stackrel{(z \rightarrow 0)}{\sim}
-
\sqrt{ \frac{\pi}{ \Theta_{l} 
\sinh  \left( \pi \Theta_{l} \right) } }
\,
\sin
\left[
\Theta_{l} \ln \left( \frac{ z }{2} \right) 
- \delta_{\Theta_{l}}
\right]
\,
\left[ 1 + O \left( z^{2} \right)  \right] 
\;  ,
\label{eq::ISP_weak_BS_asymptotic_wf}
\end{equation}
where
$ \delta_{\Theta_{l}} $ is the phase of
$\Gamma (1+i\Theta_{l})$.
In Eq.~(\ref{eq::ISP_weak_BS_asymptotic_wf}),
the wave function
oscillates with a monotonically
 increasing frequency as $r \rightarrow 0$.
As a result,
there is no criterion for the selection of a particular subset
of states and the bound-state spectrum is continuous
and not bounded from below.
Clearly, the problem should be renormalized in such a way that
the Hamiltonian recovers its self-adjoint character~\cite{sim:70}.

A first attempt~\cite{cas:50,mor:53} is to 
use Eq.~(\ref{eq::ISP_weak_BS_asymptotic_wf})
and recognize that the orthogonality 
condition for the eingenstates
restores the discrete nature of the spectrum; unfortunately,
in this approach, the Hamiltonian is not bounded from below.
However, as was proposed in Ref.~\cite{gup:93} for the particular
simple case $D=1$,
Eq.~(\ref{eq::ISP_weak_BS_asymptotic_wf})
can be regularized by introducing a short-distance cutoff
$a$, with $a \ll |E|^{-1/2}$,
so that the regular boundary condition $u_{l} (a) =0$ is implemented
in lieu of the undefined behavior at $r=0$.
Then, Eq.~(\ref{eq::ISP_weak_BS_asymptotic_wf})
gives the zeros of the modified Bessel function of the second kind
with imaginary order, $z_{n} = 2 \,
e^{ (\delta_{\Theta_{l}} - n \pi)/\Theta_{l} }$
[up to a correction factor  $1 + O(z_{n}^{2}/\Theta_{l} )$],
where $n$ is an integer; moreover,
the assumption that $z_{n} \ll 1$, with
$\Theta_{l} \geq 0$, implies that $(-n)<0$,
with the conclusion that $n= 1, 2, 3, \ldots$.
Parenthetically, $z_{n} \ll 1$ only if $\Theta_{l} \ll 1$, so that
$\delta_{\Theta_{l}}=- \gamma \Theta_{l}+ O(\Theta_{l}^{2})$
(with $\gamma$ being the Euler-Mascheroni constant) and
the energy levels become 
\begin{equation}
E_{ n_{r} l }
=
-
\left( \frac{2 \, e^{-\gamma} }{a} \right)^{2}
\,
\exp \left( - \frac{2 \pi n_{r} }{\Theta_{l}} \right)
\;  ,
\label{eq:cutoff_BS_regularized_energies}
\end{equation}
where
$n=n_{r}$ stands for the radial quantum number.

Equation~(\ref{eq:cutoff_BS_regularized_energies})
should now be renormalized by requiring that $\Theta_{l}=\Theta_{l} (a)$
in the limit $a \rightarrow 0$. More precisely,
in order for the ground state
[characterized by the quantum numbers
$
( {\rm gs} ) \equiv \left( n_{r}=1, l=0  \right)
$] 
to ``survive'' the renormalization prescription with
a finite energy, it is required that
$\Theta_{_{\rm (gs)}}    (a) 
 \stackrel{(a \rightarrow 0 )}{\longrightarrow} 
0^{+}$. This condition amounts to a
``critically strong'' coupling,
$ \lambda (a) 
 \stackrel{(a \rightarrow 0 )}{\longrightarrow} 
\lambda_{_{\rm (gs)}}^{(\ast)}
+ 0^{+}  $
(where the notation
$\Theta_{0}=\Theta_{_{\rm (gs)}}$ and
$ \lambda_{0}^{(\ast)}
 = \lambda_{_{\rm (gs)}}^{(\ast)}$
is understood for the ground state).
In particular, with this ground-state renormalization,
the required relation between $ \Theta_{_{\rm (gs)}} (a)$  
and $a$, for $a$ small,
is
\begin{equation}
- 
g^{(0)}
=
\frac{2 \, \pi}{ \Theta_{_{\rm (gs)}} (a) } +
2 \, \ln \left( \frac{\mu \, a}{2} \right)
+ 2 \, \gamma
\;  ,
\label{eq:cutoff_BS_regularized_relation}
\end{equation}
where $\mu$ is an arbitrary
renormalization scale with dimensions of inverse length
and $g^{(0)}$ is an arbitrary finite part associated with the
coupling, such that
\begin{equation}
E_{_{\rm (gs)}}
=
- \mu^{2}
\exp \left[
g^{(0)} 
\right]
\leadsto
- \mu^{2}
\;  .
\label{eq:ISP_GS}
\end{equation}
In Eq.~(\ref{eq:ISP_GS}),
 it is understood that, due to the arbitrariness of
both $g^{(0)} $
 and $\mu$, the simple choice $g^{(0)} =0$ can be made.
Finally, the ground-state wave function 
is obtained in the limit
$\Theta_{_{\rm (gs)}}    (a) 
 \stackrel{(a \rightarrow 0 )}{\longrightarrow} 
0^{+}$, which yields~\cite{gra:80}
\begin{equation}
\Psi_{_{\rm (gs)}}
 ({\bf r})
=
\sqrt{
\Gamma \left( \frac{D}{2} \right) \,
\left( 
\frac{\mu^{2}}{\pi}
\right)^{D/2}
}
\;
\frac{K_{0} (\mu r)}
{ \left( \mu r \right)^{D/2 -1} }
\;  ,
\label{eq:ISP_wf_normalized_renormalized}
\end{equation}
whose functional form, up to a factor
$ r^{-(D/2 -1)} $, is dimensionally 
invariant~\cite{gs:dimensional_invariance}.

The existence of a ground state with a dimensional scale
$\mu \propto \left| E_{_{\rm (gs)}} \right|^{1/2}$ 
violates the manifest scale invariance of the theory defined 
by Eq.~(\ref{eq:ISP_radial}), but its magnitude is totally arbitrary
and spontaneously generated by renormalization.
Here we recognize the fingerprints of 
dimensional transmutation~\cite{col:73}.

The next question refers to the possible existence 
of excited states in the renormalized theory.
For any hypothetical state with angular momentum quantum number $l >0$,
this question can be straightforwardly answered from 
the ground-state renormalization
condition
$\Theta_{_{\rm (gs)}}    (a) 
 \stackrel{(a \rightarrow 0 )}{\longrightarrow} 
0^{+}$,
which, together with Eq.~(\ref{eq:ISP_critical_coupling}),
 provides the inequality
$\lambda = 
 \lambda_{_{\rm (gs)}}^{(\ast)}
 = (D/2-1)^{2} <  \lambda_{l}^{(\ast)}$.
Then, if such a state existed, it would automatically be
pushed into the 
weak-coupling regime, with the implication that 
it could not survive the renormalization process.
This means that there are no excited states with $l>0$.
Next, the question arises as to the possible existence of bound states
with $l=0$ and $n_{r} \neq 0$.
The fact that these hypothetical bound states also cease to exist
in the renormalized theory follows from the exponential suppression
\begin{equation}
\left|
\frac{E_{n_{r} 0} }{E_{_{\rm (gs)}} } 
\right|
=
\exp \left[  - \frac{2 \pi \left( n_{r} -1 \right)
}{\Theta_{_{\rm (gs)}}}  
\right] 
 \stackrel{(\Theta_{_{\rm (gs)}}
 \rightarrow 0, n_{r} >1 )}{\longrightarrow} 
0
\;  .
\label{eq:cutoff_excited_states}
\end{equation}
Moreover, it is easy to see that, for these hypothetical states,
the corresponding limit of the wave function
becomes ill defined, so that they effectively vanish.
In conclusion, the renormalization process annihilates all
candidates for a renormalized
bound state, with the only exception of the ground state
of the regularized theory, which acquires the 
finite energy value~(\ref{eq:ISP_GS}) and the normalized
wave function~(\ref{eq:ISP_wf_normalized_renormalized}).

Similarly,
the scattering solutions can be studied by going back to
Eq.~(\ref{eq:ISP_radial}),
which implies that
$
u_{l} (r)/ 
\sqrt{r}$ is a linear combination 
of the Hankel functions $
H_{i \Theta_{l}}^{(1,2)}(kr)$~\cite{abr:72}, whose 
asymptotic behavior ($r \rightarrow \infty$), combined
with the regularized boundary condition $u_{l} (a) =0$,
provides the scattering matrix elements
$S_{l}^{(D)} (k;a)$
and phase shifts
$\delta_{l}^{(D)} (k; a)$. For example,
the phase function $\phi_{l}^{(D)} (k; a)
= \delta_{l}^{(D)} (k; a)
- ( l + D/2 -1 ) \, \pi/2 $ is given by 
\begin{equation}
\tan 
\mbox{\boldmath\Large  $\left(  \right.$ } \! \! \!  
\phi_{l}^{(D)} (k; a)
\mbox{\boldmath\Large  $\left.  \right)$ } \! \!
=
\tanh \left( \frac{\pi \Theta_{l}}{2} \right)
\,
\frac{ 1 - {\mathcal T}_{l} (k;a) \, \varrho_{l} }{ {\mathcal T}_{l} (k;a) 
+ \varrho_{l} }
\;  ,
\label{eq:ISP_cutoff_phase_shifts}
\end{equation}
where
$ {\mathcal T}_{l} (k;a) 
=
\tan \left[ \Theta_{l} \ln \left( 
ka/2 
\right) \right]$
and
$\varrho_{l}= v_{-,l}/iv_{+,l}$,
with $v_{\pm,l}=  \Gamma (1 - i \Theta_{l}) 
\pm  \Gamma (1 + i \Theta_{l}) $.
Equation~(\ref{eq:ISP_cutoff_phase_shifts})
is ill defined in the limit $a \rightarrow 0$;
in effect, the variable ${\mathcal T}_{l}(k;a)$
oscillates wildly between $-\infty$ and $\infty$,
unless $\Theta_{l} \rightarrow 0$, just as for the bound-state sector.
From Eqs.~(\ref{eq:cutoff_BS_regularized_relation})--(\ref{eq:ISP_GS}), 
when $a \rightarrow 0$, 
the renormalized $s$-wave phase shift becomes
\begin{equation}
\tan 
\mbox{\boldmath\Large  $\left(  \right.$ } \! \! \!  
\delta_{0}^{(D_{0})} (k)
 - ( D/2 - 1 )
\, \pi/2
\mbox{\boldmath\Large  $\left.  \right)$ } \! \!
=
\frac{\pi}{
\ln \left( k^{2}/|E_{_{\rm (gs)}}| \right) }
\;  .
\label{eq:ISP_cutoff_phase_shifts_renormalized}
\end{equation}
Equation~(\ref{eq:ISP_cutoff_phase_shifts_renormalized}) 
explicitly displays the scattering behavior of $s$ states, 
as well as its relation with the bound-state sector
of the theory.
Both the functional form of 
Eq.~(\ref{eq:ISP_cutoff_phase_shifts_renormalized}) and the existence
of a unique bound state in the renormalized theory are properties shared
by the two-dimensional $\delta$-function 
potential~\cite{tho:79,adh:95a,man:94}.

The analysis leading to Eq.~(\ref{eq:ISP_cutoff_phase_shifts_renormalized})
refers to $l=0$.
For all other angular momenta ($l>0$),
the coupling will be weak, so that the phase shifts 
will be given by their unregularized values, 
with the condition that $ \lambda = 
\lambda^{(\ast)}_{_{\rm (gs)}} =
\left( D/2 -1 \right)^{2}$; then,
\begin{equation}
\left.
 \delta_{l}^{(D)} 
\right|_{l \neq 0}
=
\left[
( l +  D/2 -1  )
- \sqrt{
l \, \left( l +  D -2  \right)
}
\;
\right]
 \, \frac{\pi}{2}
\; ,
\label{ISP_renormalized_phase_shifts_l_neq_0}
\end{equation}
which is a scale-invariant expression.

We now turn to an outline of a similar analysis using
dimensional renormalization~\cite{bol:72}.
In particular, we will focus on the 
bound-state sector of the theory, to
illustrate and emphasize the fact that proper renormalization
using different regularizations yields the same physics.
In this alternative regularization scheme, we 
define the dimensionally-regularized potential
in $D^{\prime}$ dimensions
in terms of its momentum-space expression,
according to~\cite{gra:80}
\begin{eqnarray} 
V^{({D^{\prime}})}( r^{\prime} )
& = &
- \lambda_{B}
\,
\int \frac{ 
    d^{{D^{\prime}}} k^{\prime} }
{(2\pi)^{{D^{\prime}}}} 
   \; e^{i{\bf k^{\prime} }  
\cdot {\bf r^{\prime} } }
\, 
\left[
\int d^{D} r \; e^{-i{\bf k} \cdot {\bf r}}
\, \frac{1}{r^{2}}
\right]_{ {\bf k} = {\bf k^{\prime} } }
\nonumber \\
& = &
- \lambda_{B}
\,
\pi^{\epsilon/2} \,
\Gamma \left(1 - \epsilon/2 \right) /
\left( r^{\prime} \right)^{ (2-\epsilon) }
\;  ,
\label{eq:fourier_ISP}
\end{eqnarray}
where $\epsilon = D-D^{\prime}$ and 
$\lambda_{B}$ is the dimensional
bare coupling, which will be rewritten as 
$\mu_{B}= \lambda \mu^{\epsilon}$, with $[\lambda]=1$
and $\mu$ being the 
floating renormalization scale.
The corresponding $D^{\prime}$-dimensional
Schr\"{o}dinger equation 
can be converted, by means of a duality 
transformation~\cite{sch:76,qui:79} 
\begin{equation} 
\left\{ \begin{array}{l} 
\left| E \right|^{1/2} r
= z^{ 2/\epsilon}  
\\ 
|E|^{-D^{\prime}/4} \, 
u_{l}(r) 
= w_{l,\epsilon} (z) 
\, z^{ 1/\epsilon -
 1/2 } 
\end{array} \right. 
\;  ,
\end{equation}
into
\begin{equation} 
\left\{
\frac{d^{2}}{dz^{2}} + 
\widetilde{\eta} 
 -
\widetilde{\mathcal V}_{\epsilon}(z) 
- \frac{ p^{2} - 1/4 }
{z^{2}}
\right\}
w_{l,\epsilon} 
(z) 
= 0
\;  ,
\label{eq:ISPSchr2}
\end{equation}
where 
$\widetilde{\mathcal V}_{\epsilon}(z) =
- 4 \, {\rm sgn} (E) \, 
z^{4/\epsilon -2}/\epsilon^{2}$.
In Eq.~(\ref{eq:ISPSchr2}), the new parameters are
\begin{equation} 
\widetilde{\eta}
 = \frac{4 \lambda 
\, \pi^{\epsilon/2} \,
\Gamma \left(1 - \epsilon/2
 \right)}{\epsilon^{2}}
\, 
\left(
\frac{|E|}{ \mu^{2} }
\right)^{- \epsilon/2 }
\;  ,
\label{eq:dual_energy_ISP}
\end{equation}
and
$
p =
2
\left(l+D^{\prime}/2 - 1 \right)/\epsilon
$.
The key to solving Eq.~(\ref{eq:ISPSchr2})
is that (i) the parameter $p$ is 
asymptotically infinite; and (ii) the
term $\widetilde{\mathcal V}_{\epsilon}(z) $
in Eq.~(\ref{eq:ISPSchr2})
behaves as an infinite
 hyperspherical potential well in the limit
 $\epsilon \rightarrow 0$.
Then, for bound states, as a first approximation,
the particle is trapped in a well with a smooth
left boundary proportional to $1/z^{2}$ and an infinite-well
boundary at $z_{2} \approx 1$; as the left
turning point is 
$z_{1} \approx p/\widetilde{\eta}^{1/2}$,
the WKB quantization condition---which we expect to be
asymptotically correct for $p \rightarrow \infty$---becomes
\begin{equation}
\int_{ p/\widetilde{\eta}^{1/2} }^{1} \sqrt{
\widetilde{\eta} 
- \frac{p^{2}-1/4}{z^{2}}  }
\; dz
\approx
\left(n_{r} -\frac{1}{4} \right) \pi
\;  ,
\label{eq:ISP_BC_WKB}
\end{equation}
so that
$
\widetilde{\eta}^{1/2}
=
p 
+
C_{n_{r}}
\,
 p^{1/3}
$,
where
$
C_{n_{r}}
=
[ 
 3 \pi \, (  n_{r} - 1/4 ) ]^{2/3}/2$.  
Therefore,
the regularized energies are
\begin{equation}
|E_{n_{r} l}|
= \mu^{2} \,
\left[
\frac{ \lambda}{  \lambda_{l}^{(\ast)} }
\right]^{2/\epsilon}
\,
\exp \left[
{\mathcal G}_{n_{r}l} (\epsilon)
\right]
\;  ,
\label{eq:ISP_DR_regularized_energies}
\end{equation}
where
\begin{equation}
{\mathcal G}_{n_{r}l} (\epsilon)
=
-2^{4/3}  C_{n_{r}} 
\left( \lambda_{l}^{(\ast)}  \right)^{-1/3} 
\epsilon^{-1/3} 
+ \left[ \ln \pi + \gamma  + 2
\left( \lambda^{(\ast)}_{l} \right)^{-1/2}
\right] 
\;  .
\end{equation}
Equation~(\ref{eq:ISP_DR_regularized_energies})
can be renormalized by 
demanding that it be finite for the ground state 
and by letting
$\lambda= \lambda (\epsilon)$; explicitly,
\begin{equation}
\lambda (\epsilon)
=
\lambda^{(\ast)}_{_{\rm (gs)}} 
\,
\left\{
1
+
 \frac{\epsilon}{2} 
\left[ g^{(0)}
- {\mathcal G}_{_{\rm (gs)}} (\epsilon)
\right]
\right\}
+
o(\epsilon)
\;  ,
\label{ISP_coupling_reg}
\end{equation}
with an arbitrary finite part $g^{(0)}$. In particular,
$\lambda (\epsilon) 
 \stackrel{(\epsilon \rightarrow 0 )}{\longrightarrow} 
\lambda^{(\ast)}_{_{\rm (gs)}}   
 + 0^{+}$, i.e., upon renormalization,
the coupling becomes critically strong with respect to $s$ states. 
Just as for cutoff regularization,
it follows that only bound states with $l=0$ 
survive the renormalization process.
As for the excited states with $l=0$
in Eq.~(\ref{eq:ISP_DR_regularized_energies}), they
are exponentially suppressed according to
\begin{equation}
\left|
\frac{E_{n_{r} 0 }  }{E_{_{\rm (gs)}} } 
\right|
=
\exp \left[ 
-2^{4/3} 
\,
 \left( C_{n_{r}} - C_{1} \right) 
\,
\left( 
\lambda^{(\ast)}_{_{\rm (gs)}}   
 \right)^{-1/3} 
\,
\epsilon^{-1/3} 
\right]
 \stackrel{(\epsilon \rightarrow 0, n_{r} >1 )}{\longrightarrow} 
0
\;  .
\label{eq:DR_excited_states}
\end{equation}

Parenthetically,
the regularized energies
of Eqs.~(\ref{eq:cutoff_BS_regularized_energies})
and (\ref{eq:ISP_DR_regularized_energies}), for finite
$a$ and $\epsilon$, are noticeably different;
nonetheless, as expected, 
their renormalized counterparts carry exactly the same 
informational content.

In short, we have analyzed
the inverse square potential and found that:
(i) a critical coupling divides the possible behaviors
into two regimes;  
(ii) 
in the strong-coupling regime, the theory 
is ill defined and requires renormalization;
and (iii) upon renormalization of the strong-coupling regime,
only one bound state survives and
$s$-wave scattering breaks scale invariance
with a characteristic logarithmic dependence.
The existence and order of magnitude 
of a critical coupling $\lambda^{(\ast)}_{_{\rm (gs)}}   =1/4$
for $D=3$ is in agreement
with recent experimental results~\cite{lev:67,des:94} 
for a wide range of polar molecules~\cite{dipole_remark}.

A final remark is in order. 
Strictly, even though a more careful treatment with dimensional 
regularization changes Eq.~(\ref{eq:ISP_DR_regularized_energies}),
the difference appears only at the level of
the finite parts (linear in $\epsilon$)
and is immaterial to the arguments presented here.
These corrections, as well as a detailed analysis of 
the scattering sector of the theory,
will be presented elsewhere.

This research was supported in part by
CONICET and ANPCyT, Argentina
(L.N.E., H.F.,
and C.A.G.C.) and by 
the University of San Francisco Faculty Development Fund
(H.E.C.).
The hospitality of the University of Houston and instructive
discussions with Prof.\ Carlos R. Ord\'{o}\~{n}ez
are gratefully acknowledged by H.E.C.


\begin{references}
\bibitem{gs:dimensional_invariance}
Incidentally, our theory in $D$ dimensions
has a two-dimensional appearance, due to dimensional
transmutation. This is in sharp contrast with the behavior
of the solutions of the wave equation 
and of the free-particle Schr\"{o}dinger equation.
See also Eq.~(\ref{eq:ISP_cutoff_phase_shifts_renormalized}).
\bibitem{dipole_remark}
The present solution of the inverse square potential (with $l=0$)
applies to the radial part of the wave function under a dipole
potential; the angular part
can be dealt with by using the method of Ref.~\cite{lev:67}.
\end{references}
\end{document}